\documentclass{article}

\usepackage[margin=1in]{geometry}
\usepackage[utf8]{inputenc}
\usepackage{amsmath}
\usepackage{amssymb}
\usepackage{amsthm}
\usepackage{xargs}
\usepackage{bm}
\usepackage{graphicx}
\usepackage{booktabs}
\usepackage{tabularx}
\usepackage[colorlinks=true, linkcolor=blue]{hyperref}
\usepackage{scalerel,stackengine}
\usepackage{xcolor}
\usepackage{multirow}
\usepackage{makecell}
\usepackage{chngcntr}
\usepackage{apptools}
\usepackage{dsfont}
\usepackage{authblk}
\usepackage[numbers]{natbib}
\usepackage{longtable}
\AtAppendix{\counterwithin{theorem}{section}}

\theoremstyle{definition}

\theoremstyle{remark}

\newcommand{\numeral}[1]{%
  \textup{\uppercase\expandafter{\romannumeral#1}}%
}

\stackMath
\newcommand\reallywidehat[1]{%
\savestack{\tmpbox}{\stretchto{%
  \scaleto{%
    \scalerel*[\widthof{\ensuremath{#1}}]{\kern.1pt\mathchar"0362\kern.1pt}%
    {\rule{0ex}{\textheight}}%WIDTH-LIMITED CIRCUMFLEX
  }{\textheight}% 
}{2.4ex}}%
\stackon[-6.9pt]{#1}{\tmpbox}%
}
\begin{document}

\newcommandx{\E}[2][1]{\mathbb E_{#1} \left[#2\right]}
\newcommandx{\V}[2][1]{\mathbb V_{#1} \left[#2\right]}
\newcommandx{\C}[2][1]{\mathbb C_{#1} \left[#2\right]}\newcommand{\iid}[0]{\overset{\text{IID}}{\sim}}
\newcommandx{\Ehat}[2][1]{\widehat{\mathbb E}_{#1} \left[#2\right]}
\newcommandx{\Vhat}[2][1]{\widehat{\mathbb V}_{#1} \left[#2\right]}
\newcommandx{\hattilde}[1]{\widehat{\widetilde{#1}}}
\newcommandx{\rightarrowp}[0]{\overset{p}{\rightarrow}}

\title{Mixed models for repeated measures should include time-by-covariate interactions to assure power gains and robustness against dropout bias relative to complete-case ANCOVA}
\author[]{Alejandro Schuler}
\date{\today}
\maketitle

\begin{abstract}
In randomized trials with continuous-valued outcomes the goal is often to estimate the difference in average outcomes between two treatment groups. However, the outcome in some trials is longitudinal, meaning that multiple measurements of the same outcome are taken over time for each subject. The target of inference in this case is often still the difference in averages at a given timepoint. One way to analyze these data is to ignore the measurements at intermediate timepoints and proceed with a standard covariate-adjusted analysis (e.g. ANCOVA) with the complete cases. However, it is generally thought that exploiting information from intermediate timepoints using mixed models for repeated measures (MMRM) a) increases power and b) more naturally ``handles'' missing data. Here we prove that neither of these conclusions is entirely correct when baseline covariates are adjusted for without including time-by-covariate interactions. We back these claims up with simulations. MMRM provides benefits over complete-cases ANCOVA in many cases, but covariate-time interaction terms should always be included to guarantee the best results. 

\end{abstract}

\section{Introduction}

Randomized trials are the gold standard for estimating causal effects because they eliminate all confounding in expectation \cite{Sox2012-oj, Overhage2013-ob, Hannan2008-sv}. However, randomized trials must still contend with random sampling variability and with bias from dropout, which cannot be decoupled from treatment by design \cite{Ashbeck2016-uz, Bell2013-zx}. These issues must be dealt with by appropriate analyses.

When the outcome of interest in the trial is continuous, e.g. a test score 12 months post-treatment, the most common definition of treatment effect is the difference between the population mean outcomes under treatment and control \cite{Chen2018-tr, Rosenblum2010-em}. The difference in sample averages between treatment arms (ANOVA) is an unbiased estimator of this effect, but suffers from high sampling variance (i.e. uncertain estimates; large p-values and confidence intervals) \cite{Wang2019-mr, Colantuoni2015-ef}. Linear models (ANCOVA) are often used to explain away variability in the outcome with baseline covariate variability and therefore increase power without introducing bias \cite{Leon2003-ee, Schuler2020-hl}. Strict type-I error control may be maintained with the use of robust standard errors \cite{Aronow2019-ms}.

Trials are often designed so that intermediate measurements of the primary outcome are collected at defined timepoints before the end of the trial \cite{Ashbeck2016-uz, Mallinckrodt2004-ov}. For example, a trial where the primary outcome is a test score at 12 months post-treatment may also dictate collection of the same test score at 3, 6, and 9 months for each subject post-treatment. These outcomes are sometimes called ``longitudinal'' or ``panel'' data. Despite the collection of these intermediate outcomes, the estimand of interest is often still the difference in mean outcomes at the final timepoint \cite{Ashbeck2016-uz, Chen2018-tr}. We call this a ``single timepoint'' effect to distinguish it from cases where the effect of interest depends on the intermediate outcomes (e.g. when the effect is the difference in slopes of the outcomes over time) \cite{Schneider2009-jw}. We also clarify that our interest here is not in a time-varying treatment, nor are we attempting to adjust for time-varying covariates.

From a statistical perspective, the stated purpose of collecting these intermediate outcomes is twofold. Firstly, having more post-treatment data should intuitively increase power. In other words, estimators that leverage this extra data should be more efficient than those that don't. Secondly, having data on intermediate timepoints should help combat the biasing effects of dropout because subjects that leave the study before the final timepoint may still contribute information in the interim.  

It is therefore common practice to employ methods that directly account for these intermediate outcomes. Mixed models for repeated measures (MMRM) are an extension of ANCOVA that are often used for this purpose \cite{Mallinckrodt2001-ay, Mallinckrod2008-ka}. 
%Generalized estimating equations (GEE) are a closely related approach that provide robust inference in the face of model misspecification \cite{}. 
We refer to MMRM as a ``longitudinal'' analysis although the target of inference is still the effect at a single timepoint. MMRM is often used with the implicit assumptions that it a) is more efficient than a single-timepoint analysis (e.g. ANCOVA) and b) protects the user from bias in a way that complete-case analysis cannot \cite{Mallinckrodt2001-ay}. 

Our purpose here is to show that both of these assumptions are generally untrue unless covariate-by-time interaction terms are included in the MMRM model specification. In fact, we prove that mathematically that MMRM can often have worse asymptotic variance (and therefore lower power) than ANCOVA if these are omitted. Our simulations confirm this fact. Moreover, we show that the missing data assumptions required for unbiased estimation using MMRM without these interactions are actually slightly more stringent than are required for complete case analysis with ANCOVA (and identical when interactions are included). This is not a novel finding, but it bears repetition since many analysts are averse to complete-case analysis in situations where they would be quite comfortable with MMRM. Using simulations, we demonstrate precisely the scenario where complete-case ANCOVA preserves type I error but MMRM without time-covariate interactions is unable to do so. 
%The story is the same for GEE and robust ANCOVA: complete case analysis with robust ANCOVA is asymptotically more efficient than GEE and unbiased under slightly more general assumptions on dropout. 

We conclude that MMRM can indeed be used to estimate single-timepoint effects, but a statistical advantage compared to complete-case ANCOVA is not assured unless time-covariate interactions are included. To our knowledge there is as of yet no explicit guidance to this effect in the literature, although many trials do include such interactions in their MMRM analyses \cite{Deodhar2016-qg, Perkovic2020-jk}. Unfortunately, many do not \cite{Muenzer_undated-kb, Whone2019-bb}. Our results are therefore of great value for trials that employ MMRM to estimate single-timepoint effects. 

\section{Estimators}

Here we present the three ways of estimating single-timepoint treatment effects from longitudinal outcomes that we will contrast for the rest of the paper. Without loss of generality we will assume that it is the treatment effect at the last timepoint that is of interest.

\paragraph{Notation}
Let $X_i = [X_{i,1}, \dots X_{i,k}, \dots X_{i,K}]^\top$ be a vector of $K$ baseline covariates for subject $i$. Presume each subject is randomized to treatment $W_i \in \{0,1\}$ with probability $\pi_1$ at the beginning of the trial. Let $Y_i = [Y_{i,1}, \dots Y_{i,j}, \dots Y_{i,J}]^\top$ be the vector of observed outcomes over time $j$ (post-randomization) for subject $i$. For any of these estimators we can also assume that there is a dropout process $T_i \in \{0 \dots J\}$ so that the outcomes for subject $i$ are observed up until timepoint $T_i$ and not after. In general $T_i$ is not independent from the baseline covariates $X_i$ or treatment $W_i$.

% \paragraph{ANOVA: Single-timepoint analysis}
% ANOVA is one of the simplest estimators used for single-timepoint analysis. The specification of the ANOVA model (for the final timepoint $J$) is

% \begin{equation}
% \begin{split}
% \label{eq:anv-model}
%     Y_{i,J} | W_i=w \sim \mathcal N(\mu_w, \sigma^2_w) \\
% \end{split}
% \end{equation}

% This model is fit using maximum likelihood. The treatment effect is then calculated using the estimate $\hat\mu_1 - \hat\mu_0$. The estimator and its model-based standard error are both consistent regardless of the true data-generating distribution for the trial. However, any dropout that is not MCAR may bias the estimate. 

\paragraph{ANCOVA}
ANCOVA is a popular estimator of the treatment effect for continuous outcomes. The specification of the ANCOVA model (for the final timepoint $J$) is

\begin{equation}
\begin{split}
\label{eq:acv-model}
    Y_{i,J} &= \alpha_J + X_i^\top\beta + W_i\tau_J + \epsilon_{i,J} \\
    \epsilon_{i,J} &\iid \mathcal{N}(0,\sigma^2_J)
\end{split}
\end{equation}

The model is fit using maximum likelihood. The coefficient $\tau_J$ is our estimate of the treatment effect at timepoint $J$. 
% It is known that this estimate converges in probability to the true effect whether or not the model is correctly specified \cite{}.
When the model is correctly specified, maximum likelihood theory provides a model-based standard error; otherwise a sandwich-type estimator may be used to perform valid inference \cite{Aronow2019-ms}.\footnote{
Throughout this paper we prioritize discussion of the model-based maximum likelihood versions of the estimators we present because these are more familiar and more often used in practice. However, we recommend using robust inference in practice in order to prevent overoptimistic standard errors when models are misspecified (as they always are). We are content to center our discussion on maximum likelihood inference because the conclusions are the same whether or not one uses robust standard errors. 
} 

Note that the ANCOVA estimator completely ignores the intermediate outcomes $Y_{i,1} \dots Y_{i,J-1}$.

\paragraph{MMRM: without time-covariate interactions}
When the outcome is longitudinal, mixed models for repeated measures (MMRM) estimation is a common approach \cite{Mallinckrodt2001-ay, Mallinckrod2008-ka}. The usual model specification of this estimator is: 

\begin{equation}
\begin{split}
\label{eq:mmrm-model}
    Y_{i,j} &= \alpha_j + X_i^\top\beta + W_i\tau_j + \epsilon_{i,j} \\
    [\epsilon_{i,1} \dots \epsilon_{i,J}]^\top &\iid \mathcal{N}(0,\Sigma)
\end{split}
\end{equation}

In the most flexible and powerful form of this estimator, the covariance matrix $\Sigma$ is left unstructured (i.e. no assumptions are placed on its form). Again this model is fit via maximum likelihood. This may be done while employing the intermediate outcomes of subjects who later drop out of the trial (see section \ref{sec:ml}). Here $\alpha_j$ and $\tau_j$ correspond to timepoint-specific intercepts and treatment effects ($\tau_J$ being the effect of interest). Another way of writing this would be to ``unravel'' the $Y_{i,j}$ into a single vector and regress onto covariates as well as time-by-intercept and time-by-treatment indicators. 

Model-based standard errors may be estimated using maximum likelihood theory if the model is presumed correct. As an alternative, a sandwich-type standard error may be estimated in order to account for misspecification of the model. MMRM with a robust standard error is equivalent to generalized estimating equations (GEE) in this context \cite{Liang1986-em}. 

Throughout the paper we will use the term ``MMRM'' to refer to this estimator.

\paragraph{MMRM$\otimes$: with time-covariate interactions} 
It is easy to extend the usual MMRM model with timepoint-specific covariate effects as follows:

\begin{equation}
\begin{split}
\label{eq:mmrm-model-interactions}
    Y_{i,j} &= \alpha_j + X_i^\top\beta_{\bm j} + W_i\tau_j + \epsilon_{i,j} \\
    [\epsilon_{i,1} \dots \epsilon_{i,J}]^\top &\iid \mathcal{N}(0,\Sigma)
\end{split}
\end{equation}

Note the $j$ subscript in the term $X_i^\top\beta_{\bm j}$. This specification is equivalent to a regression of $Y_{i,j}$ onto time-by-intercept, time-by-treatment, and time-by-covariate interaction terms.

Throughout the paper we will use the term ``MMRM$\otimes$'' to refer to this estimator.

\section{Efficiency}

\subsection{Theory}

One stated reason for using a longitudinal analysis (MMRM) over a single-timepoint analysis (ANCOVA) is that the longitudinal analysis is thought to have greater power because it incorporates the intermediate outcomes \cite{European_Medicines_Agency_undated-zp}. However, precisely the opposite is usually true unless time-covariate interaction effects are included for MMRM. Section \ref{sec:efficiency} in the appendix outlines a proof that the asymptotic variance of covariate-adjusted MMRM without time-treatment interactions is \textit{higher} or equal to that of ANCOVA when all outcomes are observed. In other words, given the same (large-sample) trial without dropout, we should expect no-interaction MMRM to produce estimates with larger confidence intervals than would be produced with ANCOVA. This translates to fewer significant findings, i.e. lower power. 
%The same result holds if robust confidence intervals are used (i.e. comparing robust ANCOVA to GEE; theorem \ref{}).

The problem with MMRM is that the adjustment term $X^\top \beta$ has to fulfil multiple roles. In the ANCOVA estimator, $\beta$ is chosen so as to minimize the prediction error for the outcome at the last timepoint, $Y_J$. In the MMRM estimator, $\beta$ must minimize the prediction error averaged across \textit{all} timepoints. Consequently, the estimate of $\beta$ resulting from the MMRM estimator ``splits the difference'' between these different timepoints and MMRM suffers in accurately predicting the outcome at the final timepoint. It turns out that this loss of predictive ability is what drives the increase in variance. Covariate adjustment functions by explaining away variance in the outcome with variance in the covariates. Since MMRM predicts worse or equally well at best, it explains away less of the variance in $Y_J$ and is therefore less efficient and has less power.

As one might expect, adding timepoint-specific linear effects for each covariate to the MMRM model specification (i.e. using MMRM$\otimes$) resolves this problem (see section \ref{sec:efficiency-interaction} in the appendix).

When dropout is accounted for the picture is more complicated mathematically and it is difficult to give a clear picture of when and why MMRM fails to be more efficient than ANCOVA. At a minimum, however, we can show that no-interaction MMRM can provide no benefit if the rate of dropout is negligible (as discussed above). When dropout is substantial and correlations exist it is difficult to mathematically delineate at which point the effects described above come to dominate and any benefits disappear. However, our task in this section is merely to demonstrate that there \textit{are} easily-understood scenarios where MMRM does fail to gain power over ANCOVA. To give a fuller characterization, we turn to simulation.

\subsection{Simulation}

Our simulations show that complete-case ANCOVA generally attains higher power than MMRM unless time-covariate interactions are included. Empirically, no-interaction MMRM can attain higher power but only if a) dropout is substantial, b) the correlation between residuals $\epsilon_{i,j}$ and $\epsilon_{i,j'}$ is high, and c) the covariate-outcome relationship is relatively stable between timepoints $1\dots J-1$ and timepoint $J$. Without any one of these factors, MMRM's advantage collapses if time-covariate interactions are not included. On the other hand, MMRM$\otimes$ dominates complete-case ANCOVA if there is any dropout and matches it exactly when there is not. These results concur with what we have shown theoretically.

% Here we present two simulations which showcase our arguments. 

% In the first, the data-generating process is such that the relationship between $X$ and $Y_j$ for the first few timepoints is very different than the relationship between $X$ and $Y_J$. As a consequence, MMRM suffers in predicting $Y_J$ and obtains substantially less power than ANCOVA (and, in fact, even less power than ANOVA). This occurs despite high inter-timepoint correlations, i.e. large values of $\Sigma_{J,j}$ for all $j$.

% In the second scenario, the relationship between $Y_j$ and $X$ is similar enough to that between $Y_J$ and $X$ that MMRM benefits from having access to the intermediate outcomes when the sample size is relatively small. As the sample size increases, however, we see that ANCOVA once again dominates. 

\subsubsection{Setup}
\label{sec:power-sims-setup}

In each simulation scenario we generated trial data according to the following data-generating process:

\begin{equation}
\begin{split}
\label{eq:power-sims}
    X_{i,k} &\sim \text{Unif}(-1,1) \ \forall \ k\in\{1\dots K\} \\
    W_{i} &\sim \text{Bernoulli}(0.5) \\
    Y_{i,j} &= \alpha_j + X_i^\top\beta_j + W_i\tau_j + \epsilon_{i,j} \\
    [\epsilon_{i,1} \dots \epsilon_{i,J}]^\top &\iid \mathcal{N}(0,\Sigma)
\end{split}
\end{equation}

Note that the effect of the covariates can change from timepoint to timepoint in this data-generating process (i.e. we have a term $X_i^\top \beta_{\bm j}$).

We used $K=2$ baseline covariates and $J=3$ timepoints. For simplicity we let $\alpha_j = 0$ and $\tau_j = 1/3$ for all $j$. We also set $\beta_j = [5, \dots 5] \in \mathbb{R}^K$ for all $j \ne J$ ($\beta_J$ is addressed later). In all cases we sampled 400 subjects into the trial (200 treated, 200 control).

We varied three parameters of the simulation. The first of these was the rate of dropout $\delta$. Outcomes were censored at a rate of $\delta$\% per timepoint. Formally, we sampled $T_i \sim \text{Geom}(\delta)$ and censored any outcomes for subject $i$ where $j \ge T_i$. Thus once a subject dropped out of the trial they could not re-enter. This missingness was independent of treatment (thus both methods produce unbiased estimates of the treatment effect; see section \ref{sec:dropout}). The second parameter of interest was the degree of correlation $\rho$ between the residuals at each timepoint. For simplicity, we simulated from an interchangeable correlation structure $\Sigma_{j,j} = 1$ and $\Sigma_{j,j'} = \rho \ (j\ne j')$. $\rho = 1$ therefore corresponds to high outcome correlation across timepoints and $\rho = 0$ to low. The last parameter of interest was the value of the coefficients at the last timepoint $\beta_J$, which we controlled via a scalar multiplier $b$ such that $\beta_J = b \beta_1$. When $b=1$ the outcome-covariate relationship is the same at the last timepoint as it is in previous timepoints but when $b \ne 1$ this relationship shifts.

We simulated data from this process 1000 times and applied the ANCOVA, MMRM, and MMRM$\otimes$ estimators to estimate treatment effects, standard errors, and p-values. For ANCOVA we used complete-case analysis using only the subjects with observed outcomes $Y_J$ whereas with MMRM and MMRM$\otimes$ we used all available outcomes data. We then calculated the power of each estimator by averaging across repetitions (significance assessed as $p< 0.05$). We repeated this entire process for different values of $\delta$ (dropout), $\rho$ (error term correlation), and $b$ (stability of covariate-outcome relationship over time).

\subsubsection{Results}

\begin{figure}[h!]
\centering
\includegraphics[width=0.75\textwidth]{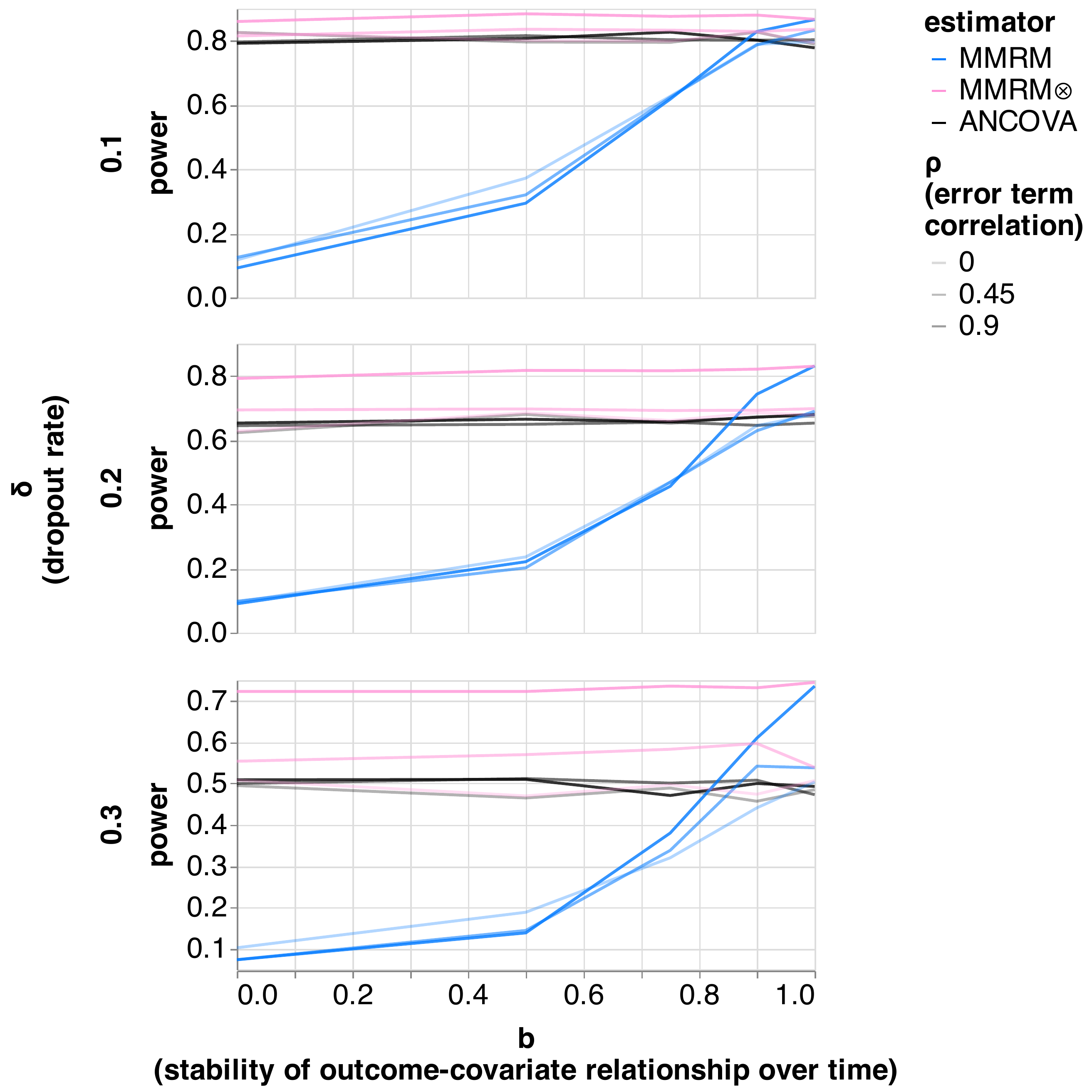}
\caption{Power of ANCOVA (complete cases), MMRM, and MMRM$\otimes$ across different simulation scenarios. All parameters ($\delta$, $\rho$, $b$) refer to the family of data-generating processes described in section \ref{sec:power-sims-setup}. Each ``point'' in the plots represents a single trial and its associated power as ascertained by simulation. The results indicate that trials analyzed with MMRM usually attain lower power than those analyzed with ANCOVA, which attain lower power than those analyzed with MMRM$\otimes$.}
\label{fig:power}
\end{figure}

Our results (figure \ref{fig:power}) show that ANCOVA generally attains higher power than MMRM. MMRM can improve power, but only under a specific constellation of parameter values. Specifically, MMRM wins when when a) dropout is substantial, b) the correlation between residuals $\epsilon_{i,j}$ and $\epsilon_{i,j'}$ is high, and c) the covariate-outcome relationship is relatively stable between timepoints $1\dots J-1$ and timepoint $J$. In all other cases ANCOVA dominates. These results align with our theoretical expectations.

Neither the dropout rate or the error term correlation appear to have a substantial impact on power when the covariate-outcome relationship is not stable over time. Even when dropout and inter-timepoint correlation is high ($\delta = 0.3$, $\rho = 0.9$), a 20\% reduction in the effect of the covariates ($b=0.8$) at the last timepoint is enough to make ANCOVA dominant. 

In scenarios where MMRM is more powerful than ANCOVA, the advantage in power is only substantial when the error term correlation is extreme ($\rho = 0.9$). This level of correlation would be considered highly unrealistic for most real trials.

On the other hand, MMRM$\otimes$ consistently dominates ANCOVA. This effect is naturally more pronounced when there is more missingness and when error term correlations are high but is not affected by shifts in the covariate effects over time.

% ADD: what about with time-covariate interactions?

\section{Sensitivity to dropout}
\label{sec:dropout}

\subsection{Theory}

MMRM is generally considered a stronger method for addressing bias from dropout than ANCOVA. This is because ANCOVA does not leverage outcomes from intermediate timepoints. Without additional modification (e.g. last value carry forward, multiple imputation), ANCOVA can only proceed by case-wise deletion of subjects whose outcome at time $J$ was not observed. MMRM, on the other hand, can still use intermediate outcomes from these subjects and therefore does not require them to be \textit{entirely} dropped from the analysis. It is somewhat intuitive that a method using intermediate data might be less prone to bias from dropout.

Unfortunately, this intuitive argument does not hold water mathematically. We show in the appendix (\ref{sec:bias}) that MMRM, MMRM$\otimes$, and complete-case ANCOVA require correct specification of the model (i.e. linearity in the covariates, errors uncorrelated with the covariates) to be insensitive to dropout when the dropout can be affected by the observed covariates and treatment (i.e. missingness is ``at random''; MAR).\footnote{When dropout is uncorrelated from the covariates (missing \textit{completely} at random; MCAR), all methods are unbiased. When dropout can depend on unmeasured covariates (missing \textit{not} at random; MNAR), no method can address potential bias without additional assumptions. Therefore only the MAR case is of interest.} The fact that MMRM requires a correct specification of the model to be unbiased in the face of MAR dropout is often glanced over in the literature \cite{Mallinckrod2008-ka}.

However, because the standard MMRM model shares the effect of covariates across timepoints, it may be misspecified in scenarios where ANCOVA and MMRM$\otimes$ model are correctly specified (i.e. when $Y_{i,J} = X_i^\top\beta + \dots $ holds but $Y_{i,j} \ne X_i^\top\beta + \dots $ for all $j$). In other words, the usual MMRM estimator may be biased by dropout if the effect of covariates varies over time, whereas complete-case ANCOVA and MMRM$\otimes$ will not be. There is no case where complete-case ANCOVA is misspecified but MMRM is not. This fact has been previously noted in the literature \cite{Dinh2011-rr}.
% The same holds for the robust alternatives (robust ANCOVA, GEE). 

Both MAR and/or linearity in the covariates are, in practice, completely unrealistic assumptions. Bias from dropout is therefore inevitable to some extent. The implication of our theoretical results is that complete-case ANCOVA and MMRM$\otimes$ are exactly unbiased when MMRM is (and in some cases when the standard MMRM model isn't). However, it should be noted that this says nothing about the relative magnitude of the bias for these estimators in cases where none of them is exactly unbiased. 

% ADD: what about when time-covariate interaction is included?
% ADD: how *bad* is the bias when both are biased?

\subsection{Simulation}

Our simulations confirm our theoretical findings. We focus on the case where there is a difference in the potential for bias, i.e. when the MMRM model is misspecified but ANCOVA and MMRM$\otimes$ are not. We find that MMRM incurs substantial type I error if the covariate-outcome relationship shifts over time, whereas complete-case ANCOVA and MMRM$\otimes$ do not.

\subsubsection{Setup}
\label{sec:error-sims-setup}

The setup for our simulations here is identical to what is described in eq. \ref{eq:power-sims}. For these simulations we set $\tau_j =0$ for all $j$ (i.e. no effect of treatment). To simulate missingness at random (conditional on treatment and covariates) we let 

\begin{align}
T_i &\sim \text{Geom}(\delta_i) \\
\delta_i &= 
\begin{cases}
    \delta + 0.1X_{i,1}  & \text{when} \ W_i = 1 \\
    \delta - 0.1X_{i,1}  & \text{when} \ W_i = 0 \\    
\end{cases}
\end{align}

% ADD: some kind of $\delta$ parameter that adjusts the overall dropout.

and censor any outcomes for which $j \ge T_i$ for subject $i$. Effectively, the time of dropout depends stochastically on the value of the first covariate and on treatment for each subject. Dropout is accelerated when subjects are treated and have a high value of $X_{i,1}$ and decelerated when the opposite holds. This represents MAR dropout.

Again we vary $\rho$ (the error term correlation; using the same interchangeable correlation structure as above), $\delta$ (the marginal dropout rate), and $b$ (the stability of the outcome-covariate relationship over time). 

We simulated data from this process 1000 times and applied the ANCOVA, MMRM, and MMRM$\otimes$ estimators to estimate treatment effects, standard errors, and p-values. For ANCOVA we used complete-case analysis using only the subjects with observed outcomes $Y_J$ whereas with MMRM and MMRM$\otimes$ we used all available outcomes data. We then calculated the type I error of each estimator by averaging across repetitions (significance assessed as $p< 0.05$). We repeated this entire process for different values of $\rho$ (error term correlation), and $b$ (stability of covariate-outcome relationship over time).

\subsubsection{Results}

\begin{figure}[h!]
\centering
\includegraphics[width=0.75\textwidth]{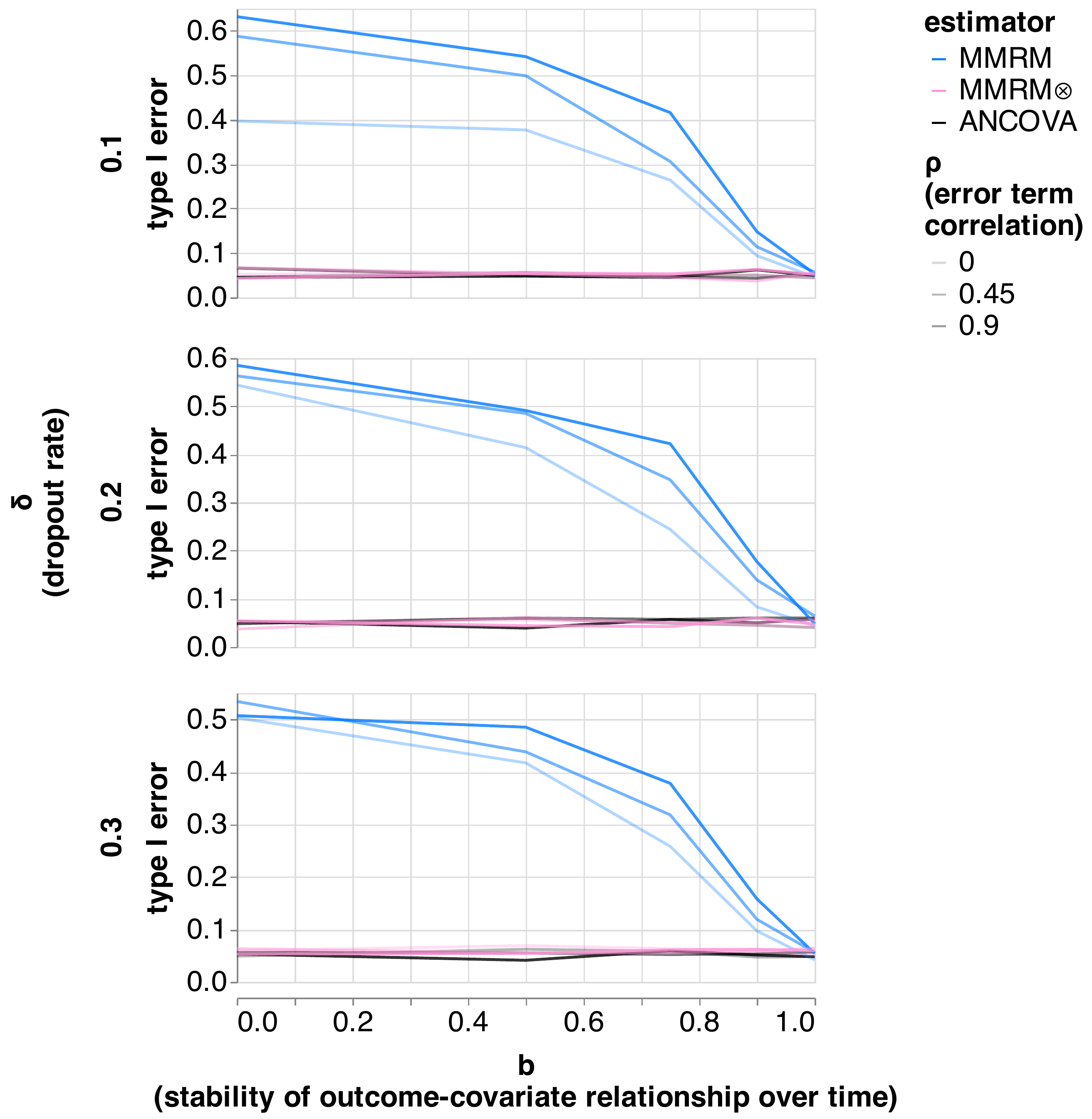}
\caption{Type I error of ANCOVA, MMRM, and MMRM$\otimes$ across different simulation scenarios. All parameters ($\delta$, $\rho$, $b$) refer to the family of data-generating processes described in section \ref{sec:error-sims-setup}. Each ``point'' in the plots represents a single trial and its associated type I error as ascertained by simulation. The results indicate that, for this class of data-generating processes, trials analyzed with MMRM typically fail to control type I error whereas those analyzed with ANCOVA or MMRM$\otimes$ succeed in doing so.}
\label{fig:error}
\end{figure}

Our results (figure \ref{fig:error}) show that the standard MMRM estimator can incur substantial type I error when the outcome-covariate relationship varies over time. On the other hand, the complete-case ANCOVA and MMRM$\otimes$ estimators maintain type I error control at the expected 5\% rate regardless of how the outcome-covariate relationship varies over time (because it is still linear at the last timepoint). The type I error of the MMRM estimator is substantial even when the effect of the covariates on the outcome is decreased by only 20\%. The inter-timepoint error correlation ($\rho$) has a relatively small effect compared to the stability of the outcome-covariate relationship ($b$), but type I error does increase with increased correlation.

\section{Discussion}

Our primary conclusions are that MMRM neither automatically increases power nor lessens sensitivity to missing data relative to complete-case ANCOVA. Adding time-covariate interactions solves the efficiency problem but strong assumptions are still required to maintain unbiasedness in the face of dropout.
% Our results pertain equally when robust standard errors are used (GEE, robust ANCOVA). 
These conclusions are supported by mathematical proof and confirmed by simulation. A number of analyses with MMRM to-date have omitted time-covariate interaction terms so our results are clearly relevant to current practice.

Of course, we do not promote the use of complete-case ANCOVA. For one, ANCOVA is not necessarily the most powerful estimator. Modern semiparametric estimators are guaranteed to improve or match its power \cite{Schuler2021-gt}. Our results also show that MMRM$\otimes$ is more powerful than ANCOVA, especially if dropout rates and inter-timepoint correlations are high.

Another argument against complete-case ANCOVA is that it relies on a well-specified linear model to remove bias due to dropout. However, it happens that this is the same assumption required by MMRM (a slightly stronger assumption, actually) and MMRM$\otimes$. In effect, performing a longitudinal analysis gives no additional protection against bias due to dropout. Instead, we encourage trialists to use theoretically grounded methods like inverse probability-of-censoring weighting (IPCW) with nonlinear regression models to ensure the analysis is as robust to dropout as possible \cite{Little2012-ko}. Sensitivity analyses should be employed to characterize the possible effects of dropout related to unmeasured covariates. Further work is necessary to characterize the magnitude of the bias incurred by MMRM and MMRM$\otimes$ vis-a-vis a complete case ANCOVA when no method is entirely successful at controlling bias.

Although our theoretical arguments cover effectively all possible data-generating processes in randomized trials, our simulations are necessarily limited to a finite number of illustrative cases. There are a few alternative scenarios that may be of particular interest to readers. 

For one, we might imagine scenarios where the effect of treatment is not constant or even monotonic in time, as is often assumed. This would do nothing to change the observed result because MMRM always includes time-by-treatment terms that allow for arbitrary changes in treatment over time (and ANCVOA is clearly oblivious to time). 

We might also consider cases where the number of covariates or timepoints is increased. Again, neither of these would fundamentally change the conclusions because the theory holds generally for all $K \ge 1$ (without covariates, MMRM and MMRM$\otimes$ are equivalent) and $J \ge 1$. In practice, increasing the number of covariates or timepoints could make it more likely that one of the covariate-outcome relationships is different at one of these timepoints from that at the final timepoint. This would make MMRM more likely to lose power and more susceptible to type I error inflation.

There are also cases where all linear-Gaussian models are misspecified (e.g. when there are ``outliers'' in the outcomes so $\epsilon$ is not normal, or when the the outcome-covariate relationship is nonlinear). One consequence of this is that all confidence intervals and p-values based on maximum likelihood estimation of linear-Gaussian models become invalid. Type I error control is therefore not guaranteed for \textit{any} of the proposed estimators, even if there is no dropout whatsoever. This problem can be resolved with the use of robust standard errors (i.e. substituting GEE for MMRM) if there is no dropout or if dropout is MCAR. However, misspecification in the face of any dropout that is not MCAR will make strict type I error control impossible for these estimators even with the use of robust standard errors because they become unable to control for bias (i.e. the standard error is correct, but the parameter is no longer causally identified). In that case, more advanced methods are required. 

As for the effect of model misspecification on power, our conclusions would be unchanged. Power would decrease for all estimators in this case (due to an increase in linearly inexplicable variance), but the relative performance of our estimators would be stable. What matters in this case is how well each estimator can approximate the best-fitting linear outcome-covariate relationship at the last timepoint, even if the true relationship is nonlinear. MMRM still has trouble here because it attempts to average over the best-fit linear relationships at all timepoints. If in addition we were to add non-MCAR dropout power could increase or decrease overall due to bias, but again the relative performance between estimators would be unaffected.

To keep things accessible we have restricted ourselves to likelihood-based analyses in this paper but our arguments pertain equally to linear estimators based on estimating equations (i.e. estimators using ``robust'' or ``sandwich'' standard errors to account for model misspecification). So, in particular, ANCOVA with robust standard errors generally enjoys a power advantage over GEE if time-covariate interactions are not included in the latter. This also raises an important point, which is that in cases where the models are misspecified our maximum likelihood theory only rigorously gives conclusions about \textit{estimated} standard errors for each estimator, not true sampling distributions. However, the analysis we describe using robust standard errors clarifies that it is indeed the true sampling variance of MMRM that inflates, not merely its maximum likelihood estimate. 

\section*{}

\subsection*{Data Sharing Statement}
{\small
Data sharing is not applicable to this article as no new data were created or analyzed in this study.}
\subsection*{Acknowledgements}
{\small
The author thanks David Walsh for helpful feedback on a previous version of this draft.}

\bibliography{references}

\appendix

\section{Additional notation and setup}

Let $W_w=1(W=w)$ and $\E{W_w}=\pi_w$ for symmetric notation. Let $Y_i^{(w)}$ denote the sequence of potential outcomes under treatment $W_i = w$. In other words, $Y_i = Y_i^{(0)}W_{0,i} + Y_i^{(1)}W_{1,i}$ Presume $(X_i, W_i, Y^{(w)}_i) \iid P(X, W, Y^{(w)})$. Note $W$ is independent from $Y^{(w)}$ and from $X$ by randomization.

The target of inference is the effect of treatment at the final timepoint (without loss of generality). Letting $\tau = \E{Y^{(1)} -Y^{(0)}} \in \mathbb R^J$ (one per timepoint), denote the effect at the final timepoint $\tau_J$.

Let $\tilde A$ represent a centered version of a random variable $A$, i.e. $A - \bar A$ or $A - \E{A}$, with use clear from context.

We begin our analysis with the MMRM model specification (eq. \ref{eq:mmrm-model}), restated here:

\begin{equation}
\begin{split}
\label{eq:mmrm-model2}
    Y_{i,j} &= \alpha_j + X_i^\top\beta + W_i\tau_j + \epsilon_{i,j} \\
    [\epsilon_{i,1} \dots \epsilon_{i,J}]^\top &\iid \mathcal{N}(0,\Sigma)
\end{split}
\end{equation}

with censoring time $T_i$ potentially dependent on $X_i$ and $W_i$ such that $T_i \perp \epsilon_i$ (i.e. outcomes are missing-at-random; MAR). Let $Y^\dagger_i = [Y_{i,1} \dots Y_{i, T_i-1}]$ be the subset of outcomes for subject $i$ up until time $T_i-1$, i.e. the outcomes that are actually observed. Let $D_{i,j} = 1(j < T_i \le j+1)$ be indicators of whether subject $i$ was last observed at time $j$. 

Let $\alpha^\top = [\alpha_1 \dots \alpha_J]$ and $\tau^\top = [\tau_1 \dots \tau_J]$ and let $\theta^\top = [\alpha^\top, \beta^\top, \tau^\top]$ so that the model parameters are $(\theta, \Sigma)$. Moreover let $Z^\top = [I, 1X^\top, WI] \in \mathcal R^{J\times (2J + K)}$ ($1$ denoting a vector of 1s of the appropriate size where clear from context) so that our model is written compactly as $Y_i = \mathcal N(Z_i^\top \theta, \Sigma)$.

Note that ANCOVA is a special case of this model when $J = 1$. 

\section{Maximum likelihood}
\label{sec:ml}

Under standard maximum likelihood theory, we proceed my maximizing the (log) likelihood of the parameters $(\theta, \Sigma)$ given the observables $(Y^\dagger, T, X, W)$. Under our assumptions the joint density of observables factors as $P(Y^\dagger, T, X, W) = P(Y^\dagger | X, W)P(T | X, W)P(X,W)$ so maximizing the log-likelihood $l(\theta, \Sigma; Y^\dagger | X, W)$ is all that is necessary because the other components of the joint likelihood are unaffected by the parameters.

Omitting $i$ subscripts, the expression for the derivative of the empirical log-likelihood w.r.t. $\theta$ is 

\begin{equation}
    \frac{\partial l}{\partial \theta} 
    = 
    \sum_i Z \Omega (Y-Z^\top \theta)
\end{equation}

where 
\begin{align}
    \Omega(\Sigma) &= \sum_j D_j \Omega_j(\Sigma) \\
    \Omega_j(\Sigma) &=
    \left[
    \begin{array}{cc}
    \Sigma_{[1:j, 1:j]}^{-1} & 0 \\
    0 & 0
    \end{array}
    \right]
\end{align}

If all outcomes are observed for all subjects (i.e. $Y^\dagger = Y$) then $\Omega(\Sigma) = \Sigma^{-1}$. If only outcomes up to time $j$ are observed for subject $i$, however, then the contribution to the log-likelihood comes only from the distribution of the outcomes up until the point of censoring, which creates a term like $Z_{[:, 1:j]} (\Sigma_{[1:j, 1:j]})^{-1} (Y^\dagger - (Z_{[:, 1:j]})^\top \theta)$. We use the indicators $D_j$ and a little matrix algebra to rewrite the sum of these terms in the form shown above. To our knowledge, this is the first explicit construction of the MMRM estimator with missing data that is available in the literature. For the ANCOVA estimator, $\Omega$ reduces to $D_J$.

Let $\mathcal C_{jj'}$ be the set of subjects who have their outcomes observed at times $j$ and $j'$, i.e. $\mathcal C_{jj'} = \{i:j,j' < T_i\}$. Let $R(\theta) = Y-Z^\top\theta$ be the vector of residuals. Our estimates $(\hat\theta, \hat\Sigma)$ are therefore the solution to the following score equations

\begin{align}
    0
    &= 
    \sum_i Z \Omega(\hat\Sigma) (Y-Z^\top \hat\theta) 
    \\
    0 &= \hat\Sigma_{jj'} - \frac{1}{|\mathcal C_{jj'}|} \sum_{i \in \mathcal C_{jj'}} R_j(\hat\theta)  R_{j'}(\hat\theta)
\end{align}

Without any assumptions on the veracity of the model, the theory of M-estimation guarantees us that $(\hat\theta, \hat\Sigma) \rightarrowp (\theta^*, \Sigma^*)$ where, in particular, $\theta^* = \E{Z\Omega(\Sigma^*)Z^\top}^{-1}\E{Z\Omega(\Sigma^*)Y}$ \cite{Stefanski2002-zz}.

\subsection{Efficiency}

Under standard maximum likelihood theory, the asymptotic distribution of $\sqrt{n}(\hat\theta - \theta)$ is 
$\mathcal N(0, I(\theta)^{-1})$ where $I(\theta)$ is the Fisher information of $\theta$. In this case the Fisher information is

\begin{align}
    I(\theta) 
    & = 
    \E{Z\Omega Z^\top} 
    \\
    &= 
    \sum_j \E{Z\Omega_jZ^\top D_j} 
    \\
    & =
    \left[ 
    \begin{array}{ccc}
         \sum_j \Omega_j d_j & 0 & 0 \\
         0 & \sum_j d_j \omega_j V_j & \sum_j d_j p_{0,j} p_{1,j} \Delta_j 1^\top \Omega_j \\
         0 & \sum_j d_j p_{0,j} p_{1,j} \Omega_j 1\Delta_j^\top & \sum_j d_j \Omega_j p_{0,j} p_{1,j}
    \end{array}
    \right]
\end{align}

We have introduced a large number of new terms, which we define here:

\begin{align*}
    d_j &= \E{D_j} \\
    \omega_j &= 1^\top \Omega_j 1 \\
    V_j &= \V{X|D_j} \\
    p_{w,j} &= \E{W_w | D_j} \\
    \Delta_j &= \E{X|D_j, W=1} - \E{X|D_j, W=0}
\end{align*}

The key term here is $\Delta_j$, which can be understood as a measurement of the confounding induced by the dropout at time $j$.

To obtain this result we use the definitions provided above and basic properties of conditional expectation, etc. The only ``clever trick'' that is required is a centering scheme that substitutes $\tilde X_i = \sum_j D_{i,j} (X_i - \E{X|D_j})$ and $\tilde W_i = \sum_j D_{i,j} (W_i - \E{W|D_j})$ for $X_i$ and $W_i$ in the computation. Note that this centering is feasible in practice because the means that are required are simply the empirical means of each variable among subjects who were observed until a given timepoint. Regardless, the centering does not affect the large-sample properties of the estimator, but substantially simplifies computation. 

Given this expression for the Fisher information, we can come to the conclusions discussed in the body of the paper. Naturally, the true variance of the estimated treatment effect will not be the same as the inverse Fisher information if the model was not well specified. However, the \textit{estimate} of the Fisher information (which determines power if that estimator is used) still converges to a related asymptotic limit, regardless of whether or not it still accurately represents the true sampling variance of the estimator.

Our subsequent theory focuses on the much simpler case where there is no dropout because the resulting Fisher information is more tractable. We provide the more general Fisher information here as a contribution for future work.

\subsubsection{ANCOVA has higher power than MMRM when there is no dropout.}
\label{sec:efficiency}
First and foremost, we show how the MMRM estimator can only be as efficient as or less efficient than ANCOVA when all outcomes are measured. In this case our expression for the Fisher information simplifies because $\Omega = \Sigma^{-1}$ so we have 

\begin{align}
    I(\theta) 
    & = 
    \E{Z\Sigma^{-1} Z^\top} 
    \\
    &= 
    \left[
    \begin{array}{ccc}
        \Sigma^{-1} & 0 & 0 \\
        0 & sV & 0 \\
        0 & 0 & \pi_0\pi_1 \Sigma^{-1}  \\
    \end{array}
    \right]
\end{align}

where $s = 1^\top \Sigma^{-1} 1$ and $V=\V{X}$. We are interested in the bottom-right term in the inverse of this matrix since that corresponds to the asymptotic limit of the estimated variance of $\hat\tau_J$. Proceeding with block inversion we see that $\sqrt{n}(\hat\tau - \tau) \rightsquigarrow \mathcal N(0, (\pi_0\pi_1)^{-1}\Sigma)$. The asymptotic estimated covariance matrix of the treatment effects therefore depends entirely on the asymptotic limit of $\Sigma$, which is $\hat\Sigma \rightarrowp \Sigma^* = \E{RR^T}$ with $R = R(\theta^*) = Y-Z\theta^*$ being the asymptotic residuals of the MMRM estimator (i.e. $\hat\theta \rightarrowp \theta^*$ for MMRM). We have, therefore, that $\sqrt{n}(\hat\tau_J - \tau_J) \rightsquigarrow \mathcal N(0, (\pi_0\pi_1)^{-1}\E{R_J(\theta^*)^2})$. 

Let $\theta^{**}$ denote the asymptotic limit of the parameter estimates from the ANCOVA estimator. Then, for ANCOVA, the result is simply $\sqrt{n}(\hat\tau_J - \tau_J) \rightsquigarrow \mathcal N(0, (\pi_0\pi_1)^{-1}\E{R_J(\theta^{**})^2})$ because $\Sigma = \sigma^2$.

We have now reached the crux of our argument. The only difference in the asymptotic limit of the estimated variance of MMRM and that of ANCOVA depends on the relative values of $\E{R_J(\theta^*)^2}$ and $\E{R_J(\theta^{**})^2}$. These are nothing but the mean-squared prediction errors for $Y_J$ of both estimators. Recall $R_J(\theta) = Y_J - (\alpha_J + X^\top \beta + W\tau_J)$. Approaching this as a pure optimization problem (and knowing $X \perp W$), the parameters that minimize $\E{R_J(\theta)^2}$ are $\alpha_J = \E{Y_J | W=0}$, $\beta = \V{X}^{-1}\C{X,Y_J}$ and $\tau = \E{Y|W=1} - \E{Y|W=0}$. This is a consequence of least-squares optimization.

Using the aforementioned standard result relating to the consistency of M-estimators (or even OLS) it can be shown that $\theta^{**}$ takes precisely the optimal values given above, regardless of whether or not the specification of the model is correct. Consequently, the asymptotic limit of the estimated variance of the ANCOVA estimator is less than or equal to the asymptotic limit of the estimated variance of the MMRM estimator. 

The problem with the MMRM estimator is that $\beta^* = \V{X}^{-1} \C{X,Y} \Sigma^{*-1} 1 s^{-1}$ ($s$ here is $1^\top \Sigma^{*-1} 1$). This result again follows from the usual M-estimator theory. If the assumed model (eq. \ref{eq:mmrm-model2}) does not hold, then it is unlikely that $\C{X,Y} \Sigma^{*-1} 1 s^{-1} = \C{X, Y_J}$. Of course, if that model does indeed hold, then $\C{X,Y} = \C{X,Y_J} 1^\top$ and the exact equivalence follows (i.e. MMRM and ANCOVA are equally efficient).

The implications for power follow immediately. By a continuity argument in $d_j$, any power advantage that MMRM has over ANCOVA when dropout is present must disappear as the rate of dropout approaches zero.

\subsubsection{Including covariate-time interaction terms resolves the problem.} 
\label{sec:efficiency-interaction}
Consider a variant of our MMRM estimator with $Z = [I, I \otimes \tilde X, \tilde W I]$ ($\otimes$ is the Kronecker product). This corresponds to a model with interactions between baseline covariates and timepoint indicators. Let $\beta = [\beta_1^\top \dots \beta_J^\top]^\top$ with $\beta_j \in \mathbb R^K$. Then $\beta_j \rightarrowp \beta_j^{*} =  \V{X}^{-1}\C{X,Y_j}$. The asymptotic estimated variance of the treatment effect $\tau_J$ remains $(\pi_0\pi_1)^{-1}\E{R_J(\theta^{*})^2}$. MMRM is therefore equally as efficient as ANCOVA because $\theta^{**}$ is still the unique optimizer and $\theta^{*} = \theta^{**}$.

The proof of this again follows from standard M-estimator theory. When evaluating $\E{Z\Sigma^{-1} Z^\top}$, instead of a term 
$\E{\tilde X1^\top \Sigma^{-1} 1\tilde X^\top}$
we find ourselves with 
$
\E{
(I\otimes \tilde X) 
\Sigma^{-1} 
(I \otimes \tilde X)^\top
}
$, which comes out to be 
$
\Sigma^{-1}
\otimes
\E{\tilde X \tilde X^\top}
$
after some matrix algebra. Similarly, when computing $\E{Z\Sigma^{-1}Y}$ we obtain a term 
$\E{(I\otimes \tilde X) \Sigma^{-1} Y}$
in place of 
$\E{\tilde X 1^\top \Sigma^{-1} Y}$. This term evaluates to 
$
[
(\xi(\Sigma^{-1})_1)^\top
\dots
(\xi(\Sigma^{-1})_J)^\top
]^\top
$
where $\xi = \C{X,Y}$ and we let $(\Sigma^{-1})_j$ be the $j$th column of $\Sigma^{-1}$. We therefore obtain $\beta^*_j = \V{X}^{-1}\C{X,Y}\Sigma(\Sigma^{-1})_j = \V{X}^{-1}\C{X,Y_j}$.

\subsection{Bias in the face of dropout}
\label{sec:bias}

In our maximum likelihood formulation the dropout mechanism does not impact the estimation of the model parameters as long as $T \perp \epsilon$. By standard theory, the estimated parameters converge to their true values when the model is well-specified. This proves that the MMRM, MMRM$\otimes$, and ANCOVA estimates are all still consistent for $\tau_J$ when dropout is MAR \textit{and} the model is well-specified. By definition, ANCOVA and MMRM$\otimes$ are well-specified in all cases that MMRM is. The same fact can be proven using the M-estimator consistency result discussed above, plugging in the relevant model specification (e.g. \ref{eq:mmrm-model2}) where required to show that $\hat\tau \rightarrowp \tau$.

\end{document}